\documentclass[10pt,conference]{IEEEtran}
\usepackage[utf8]{inputenc}
\UseRawInputEncoding
\usepackage[numbers]{natbib}
\usepackage{setspace}
\usepackage{graphicx}
\usepackage{amsmath, amssymb}
\usepackage{hyperref}
\usepackage{xcolor}
\usepackage{cleveref}
\usepackage{todonotes}
\usepackage{enumitem}

\begin{document}

\title{Introducing Quantum Computing to High-School Curricula: A Global Perspective} 

\author{
  \IEEEauthorblockN{
    Maria Gragera-Garces\IEEEauthorrefmark{1}\IEEEauthorrefmark{2},
    Luis Gomez-Orzechowski\IEEEauthorrefmark{1}\IEEEauthorrefmark{4},
    Juan F. Rodriguez-Hernandez\IEEEauthorrefmark{1}\IEEEauthorrefmark{3}
  }
  \IEEEauthorblockA{
    \IEEEauthorrefmark{2} University of Edinburgh\\
    \IEEEauthorrefmark{4} Universidad Autonoma de Madrid\\
    \IEEEauthorrefmark{3} Kipu Quantum GmbH\\
    \IEEEauthorrefmark{1} bqb Quantum Youth
  }
}

\maketitle

\begin{abstract}
Quantum computing is an emerging field with growing implications across science and industry, making early educational exposure increasingly important. This paper examines how quantum computing concepts can be introduced into high-school STEM curricula within existing structures to enhance foundational learning in mathematics, computer science, and physics. We outline a modular integration strategy introducing key quantum ideas into standard courses, leveraging open-source educational resources to ensure global accessibility. Emphasis is placed on educational opportunity and equity: the approach is designed to be inclusive and to bridge current curricular gaps so that students worldwide can develop basic quantum literacy. Our analysis demonstrates that integrating quantum topics at the secondary level is feasible and can enrich STEM learning.
\end{abstract}


\section{Introduction} \label{sec:intro}

Quantum computing is attracting increasing attention as a promising area for scientific and technological development. Although its practical applications are still emerging, governments, industries, and academic institutions are beginning to invest in building the foundations of a future quantum-ready
workforce \cite{greinertAdvancingQuantumTechnology2024corr, preskillNISQMegaquopMachine2025corr}. 
However, despite this growing interest, quantum computing remains largely absent from high-school curricula, hindering efforts to prepare the next generation of researchers.

Several countries have taken initial steps to incorporate quantum-related content into their education systems \cite{kaurDefiningQuantumWorkforce2022corr}. 
These efforts range from national curriculum revisions to localized pilot programs and interdisciplinary student projects \cite{bondaniIntroducingQuantumTechnologies2022corr, internationalbaccalaureateInterdisciplinaryTeachingLearning2021_1}. 
However, access to quantum learning opportunities remains uneven \cite{meyerDisparitiesAccessUS2024, tenjo-patinoQuantumComputingEducation2025}. Many students, especially in under-resourced regions, are unaware that the field of quantum computing exists. Those who do are often met with a lack of structured opportunities to explore quantum topics before reaching higher education. This gap underscores the need for a broader approach to quantum education that can be adopted in various educational systems.

Early exposure to foundational ideas can support broader goals in science education. 
These include promoting scientific literacy \cite{eshachShouldScienceBe2005corr}, fostering critical thinking and informed decision-making on science-related issues \cite{kelpDevelopingScienceLiteracy2023corr, sjogrenConnectingEthicalReasoning2023corr}, and encouraging engagement with contemporary research. In this way, early quantum education aligns with the existing objectives of secondary STEM curricula, complementing standard learning outcomes.

In this paper, we investigate practical ways to integrate quantum computing into existing high-school STEM curricula, specifically within mathematics, physics, and computer science courses. We identify key topics already present in standard curricula that naturally connect to quantum computing concepts, and we highlight notable gaps where new content could be introduced. Rather than proposing a standalone quantum computing course, we suggest a modular approach: quantum concepts are included in subjects that students are already learning, using hands-on projects and freely available materials developed by the research community.

The remainder of the paper is structured as follows. \Cref{sec:curricula} reviews current curricula in multiple countries, identifying relevant topics and gaps to incorporate quantum education.
\Cref{sec:global-initiatives} discusses the role of open-source materials in supporting accessible quantum learning and addresses challenges such as content quality and teacher training.
\Cref{sec:open-source} concludes with recommendations for implementation, emphasizing strategies to achieve inclusive and sustainable quantum literacy on a global scale.

\section{Quantum Topics and Gaps in Modern High-School Education} \label{sec:curricula}

This section aims to explore existing high-school curricula and to identify subject areas and project-based opportunities where quantum computing concepts could be meaningfully introduced. Recognizing that curricula are already saturated \cite{oecd2022curriculum},
and that quantum computing remains a highly specialized topic, we do not propose it as a standalone subject. Instead, our goal is to pinpoint core subjects and existing project formats across different countries that naturally align with quantum principles. We highlight why these areas are relevant to quantum computing and identify some gaps that, if addressed, could make integration more effective and feasible.

\subsection{Mathematics}
Mathematics provides the language to describe quantum states, operations, and behaviors. Strong knowledge in relevant mathematical areas would enable students to better understand the underlying mechanisms of quantum computing and prepare them to engage with this field at a deeper level. We identify three key areas of mathematics commonly found in high-school curricula:

\begin{enumerate}
    \item \emph{Linear Algebra}: provides the language for quantum states, gates, and algorithms. Without familiarity with vectors and matrices, students will not understand how quantum operations work.

    \item \emph{Probability Theory}: underpins the probabilistic nature of quantum mechanics. Measurement results are governed by probability amplitudes, hence students need a strong understanding of probability rules.

    \item \emph{Complex Numbers}: are essential for representing the wave-particle duality of quantum systems, such as quantum amplitudes, superpositions, and phase relationships.
\end{enumerate}

Although these topics appear in many national curricula, the depth and consistency of coverage vary widely. In the United States, for example, linear algebra and probability are typically introduced in advanced courses such as precalculus or statistics, which means that many students may never encounter them in high school \cite{commoncorestatestandardsinitiativeCommonCoreState2022corr}. In South Africa, the curriculum includes limited exposure to algebra and probability and omits complex numbers entirely \cite{departmentofeducationNationalCurriculumStatement2006}. This reflects a broader trend in which complex numbers are treated as optional or advanced content in many educational systems. 

Research suggests that structured, standards-based curricula can provide students with a more comprehensive mathematical foundation, potentially leading to better preparedness for advanced studies \cite{shafer2013impact}. The International Baccalaureate (IB) Mathematics: Analysis and Approaches Higher Level (AA HL) \cite{InternationalBaccalaureateDiploma2021corr,clastifyIBMathsAAcorr} program exemplifies this approach by including all three areas as integral parts of its core curriculum. 

Another challenge in mathematics education is the disconnect students often feel between abstract concepts and their practical use. Research shows that students report higher interest and motivation when they perceive classroom content as relevant to real-world contexts \cite{applied_maths}. Educators can enhance conceptual understanding and foster engagement by demonstrating how mathematical concepts underpin transformative technologies in quantum computing.

\subsection{Physics}
Physics curricula predominantly focus on concepts developed in the 17th and 18th centuries, with minimal inclusion of modern physics topics \cite{Costa1995}.
This poses a challenge for introducing emerging fields like quantum computing. 
However, several areas already present in existing physics curricula provide promising entry points for meaningful integration:

\begin{enumerate}
    \item \emph{Foundations of Quantum Mechanics}: introduces core principles such as superposition and entanglement.
    These concepts help students understand how quantum systems enable novel forms of computation that can solve problems previously considered intractable.

    \item \emph{Material Science}: provides the basis for understanding how quantum devices are physically built.  
    From specialized materials to control parameters of quantum interactions, this knowledge helps students grasp the hardware constraints behind qubit technologies.
    
    \item \emph{Thermodynamics}: is essential for understanding the environmental conditions required by quantum computers.  
    Stable temperatures and low electromagnetic noise are crucial for maintaining coherence and ensuring reliable system behavior.

\end{enumerate}

Since the early 2000s, countries such as Portugal, Spain, France, Italy, the United Kingdom, Denmark, Sweden, Finland, Australia, and Canada have gradually integrated quantum concepts into their high-school physics programs to varying extents \cite{stadermannAnalysisSecondarySchool2019corr}. 
However, implementation remains uneven \cite{Stadermann_2021}. Ongoing debates question how to effectively introduce quantum mechanics or if it should even be introduced at this level. Scholars warn that poor analogies and misconceptions can hinder student understanding and foster long-term confusion \cite{azamAdvancedLevelPhysics1995, TeachingModernPhysicsmisconceptions}.

In contrast, subjects like material science and thermodynamics are more consistently established in high-school curricula.
Still, significant disparities exist globally. It has been shown that different curricula 
can significantly impact students’ achievements and problem-solving skills in physics \cite{disparities}. 
These studies collectively suggest a need for a more standardized and comprehensive integration of modern physics into high-school education. 

Another common issue in high-school physics education is the heavy reliance on formula-based teaching, where students are primarily taught to apply equations without fully understanding the underlying concepts \cite{Kanderakis2016}.
While this method can help students solve standard problems, it often fails to support deeper learning, especially in abstract areas like quantum mechanics. 
Research shows that students gain a better understanding when mathematical content is taught alongside visual aids and real-world examples. These approaches help students see the meaning behind the formulas, making the subject more accessible and engaging \cite{Conana2024}.

\subsection{Computer Science}
Computer Science (CS) is a relatively new subject in high-school education and, as a result, is not yet globally adopted. 
Despite growing demand for digital skills and the expanding role of technology in society \cite{RegulationFourthIndustrialcorr}, many countries have yet to integrate CS into their high-school curricula \cite{Vegas2021Buildingcorr}. 

Amongst the countries that offer this module, implementation varies significantly. 
The United States was an early adopter, with the state of Illinois introducing a high-school-level CS course as early as the 1960s. 
In contrast, countries such as Brazil \cite{ribeiroBrazilianSchoolComputing2023corr} and Chile \cite{fowlerHowChileImplemented} have only integrated it into their high-school curricula within the last decade.

A key gap in current CS education is its focus on classical models of information processing. 
While this forms an essential foundation, it does not introduce students to alternative ways of computation. To support this integration, we identify three core areas that are relevant to quantum computing:

\begin{enumerate}
    \item \emph{Fundamentals of Computer Systems}: provides the foundation for understanding how classical devices operate, from basic logic gates to system architectures and computational complexity.
    This helps students distinguish classical systems from quantum ones in terms of how they process information.

    \item \emph{Programming Skills}: introduces students to coding principles and algorithmic thinking.  
    These skills are essential for exploring quantum programs.
    
    \item \emph{Data Representation}: explains how information is stored and manipulated in digital systems.  
    Understanding binary digits (bits) versus quantum bits (qubits) is key to grasping the differences between classical and quantum information.
\end{enumerate} 

Integrating quantum information concepts into CS courses can broaden students’ understanding of what information is and how it can be manipulated within different computational frameworks. Research has shown that introducing foundational quantum concepts through scaffolded instruction and analogies grounded in classical computation helps bridge the conceptual gap between classical and quantum models, especially for students without prior exposure to quantum mechanics \cite{mjeda2024quantumcomputingeducationcomputer}. These analogies enable learners to develop a more complete understanding of the distinct strengths and limitations of each model. In turn, this prepares them to select appropriate tools for different types of problems and equips them with the foundational knowledge needed to engage meaningfully with quantum computing.

\subsection{Quantum projects: tested drivers of interest}
\label{sec:projects}

Project-based learning is already a common part of many high-school systems, typically through activities like extended essays, research projects, and student-led investigations \cite{thomas2000review}. These formats offer another practical way to introduce quantum computing concepts without requiring major changes to existing curricula. This subsection looks at how quantum topics have been successfully included in such projects, focusing on examples from the  International Baccalaureate (IB) and the Spanish Excellence Baccalaureate (SEB).

The IB Extended Essay is a research project where students independently explore a topic of their choice. In recent years, students have used this format to investigate quantum computing, for instance, in simulating quantum search algorithms \cite{nazMENCIONHONORPROFESORES}, or exploring other quantum computing concepts \cite{Stepintothefuture}.

One notable example comes from the second author of this paper, who, as part of his SEB project, developed a visual and interactive module to teach the BB84 Quantum Key Distribution protocol to high-school students. This project received an honorary mention from Saint Louis University in Madrid and helped spark his interest in quantum computing \cite{SaintLouisUniversityLuisGomez}.

These experiences suggest that high-school students can engage meaningfully with quantum topics, even if they are complex. Research also shows that these projects help students develop valuable academic skills that go beyond traditional classroom learning. These include critical thinking, data analysis, and increased confidence in tackling university-level material \cite{inkelas2012IB, wray2013IBperception}.

However, there are still some obstacles. One major challenge is that many teachers and project supervisors are not familiar with quantum computing themselves. Without proper training or resources, it can be difficult for them to efficiently support students \cite{wray2013IBperception, lafuente2019pbl, pelaez2013exploring}. Addressing this issue will require new professional development programs and resources designed specifically for educators.

Another barrier is that most students are unaware of quantum computing as a possible project topic. If the subject has not been introduced earlier in their education, they are unlikely to choose it for a self-directed research task \cite{pelaez2013exploring}. 
Integrating the topics presented in the previous subsections into the curriculum, or guiding students through the openly available resources discussed in \Cref{sec:open-source}, can help them become more familiar with the field and spark interest in researching these topics.

\section{Existing Quantum Education Global Initiatives} \label{sec:global-initiatives}
In this section, we explore the integration of quantum computing in high-school curricula through various examples.
These efforts are gaining traction worldwide, driven by various national strategies and educational reforms.
It is important to note that the countries and consortia listed below make up a significant proportion of the global quantum investment. 
As such, they often account for quantum technologies as one of the key areas of focus in their national strategies.

\subsection{United States}
In the United States, the National Science Foundation (NSF) and the White House Office of Science and Technology Policy have launched the National Q-12 Education Partnership \cite{NSF_Q12corr}. 
This initiative aims to expand quantum education directly to middle- and high-school students by providing resources and training to educators through the Q2Work Program \cite{NSF_Award_2039745corr}, 
preparing a future quantum-ready workforce \cite{NSF_Q12corr}. 
Additionally, workshops and discussions have been conducted to explore the establishment of a National Center for Quantum Education, 
focusing on enhancing the education of quantum information science at the K-12 level \cite{Barnes2024corr}.

\subsection{United Kingdom}
The United Kingdom government has articulated a comprehensive National Quantum Strategy, outlining a ten-year vision to position the country as a leader in quantum technologies. 
A significant aspect of this strategy is the emphasis on education and skills development, including the integration of quantum concepts into the educational system. 
The strategy highlights the necessity of developing a diverse and well-rounded workforce adept in quantum technologies \cite{UK_National_Strategycorr}. 
Furthermore, more than £1.1 billion have been allocated to train thousands in future technologies, including quantum computing \cite{UK_Skills_Investmentcorr}.
While most of this funding is assigned to higher education and research, some has been allocated to summer schools \cite{NQCCQuantumExperiencecorr} and work experiences \cite{ApplyRALcorr} have been organized for secondary-education students in quantum.

\subsection{European Union}
The European Union, through the Quantum Flagship initiative, has established the Education Coordination and Support Action (QTEdu)~\cite{QTEducorr}. 
QTEdu aims to create a learning ecosystem to inform and educate society on quantum technologies. 
This includes developing teaching materials, conducting pilot projects, and fostering collaboration among educators, researchers, and policymakers to integrate quantum concepts into secondary education.
As of 2021, QTEdu has been running two relevant pilots for high-school quantum education: PHONQEE and PCK.
PHONQEE uses games and hands-on experiments to make quantum physics engaging and concrete for students, 
while PCK maps core quantum concepts to effective teaching strategies, supporting educators with a practical classroom guide.

\subsection{Canada} 
Canada is advancing quantum education at the high-school level through both institutional and government efforts. 
The Institute for Quantum Computing (IQC) at the University of Waterloo leads initiatives like Quantum Educators \cite{QuantumEducatorsInstitutecorr}, which offers professional development and open resources to help integrate quantum topics into existing curricula. 
IQC also developed QuEST \cite{IQCQuEST2022}, a toolkit with hands-on materials for teaching quantum mechanics in schools. 
Nationally, the 2023 National Quantum Strategy \cite{CanadaNQS2023} allocates \$360 million to support research, talent development, and education, emphasizing the need to build a strong quantum-ready workforce at all levels.

\section{The Role of Open Source in Global Quantum Education} \label{sec:open-source}

In this section, we explore the role of open source as both a driver and enabler of global quantum education at the high-school level. While \cref{sec:curricula} examined how quantum science can be introduced through formal curricula, open-source ecosystems offer an alternative and highly complementary route: one that supports both classroom-based and self-directed learning. This dual approach is especially valuable in contexts where educational systems face limitations in access to up-to-date materials, trained instructors, or institutional support.

We refer to open source as the collection of freely accessible tools, content, and platforms that anyone can use, modify, or share for teaching and learning quantum science. This includes open-source software, open-access educational materials such as textbooks and curricula released under Creative Commons licenses, publicly available research papers and tutorials, and community-driven initiatives.

A well-known example is IBM’s Qiskit, an open-source quantum computing software development kit (SDK) accompanied by interactive courses and tutorials that demonstrate its applications \cite{IBMQuantumLearning}. These resources are freely available online and have been translated into several languages, making them accessible to a global audience of learners and educators.

\subsection{Pilots and Practices in Open Quantum Education} \label{sec:testedmat}

A growing number of pilot programs, curriculum initiatives, and educational studies have tested how quantum computing can be introduced effectively in high-school settings \cite{PilotingFullyearOpticsbased, perryQuantumComputingHigh2020corr, sunComputingQuantumMechanics2024corr, bondaniIntroducingQuantumTechnologies2022corr}. 
These efforts demonstrate that students can engage with quantum concepts meaningfully and explore diverse instructional formats that enhance learning and accessibility.

In this subsection, we highlight several formats and tools emerging from these initiatives, 
including pictorial learning, interactive games, hackathons, and summer schools, which have shown promise in promoting student understanding and motivation.

\subsubsection*{Pictorial learning} A recurring feature across successful programs is the use of well-designed visual content to convey abstract quantum ideas intuitively. 
Quantum pictorialism, for example, has helped high-school students develop conceptual understanding without relying on advanced mathematical formalism. 
In one case study, students aged 16–18 across the United Kingdom received 16 hours of instruction using this method. 
They later sat a graduate-level quantum computing exam, with 82\% passing and nearly half earning distinction \cite{dundar-coeckeQuantumPicturalismLearning2023corr}.

\subsubsection*{Videogames} Diagrammatic approaches are often complemented by interactive platforms with game-like interfaces. 
Tools such as Quarks Interactive visualize principles like superposition, entanglement, and interference through interactive simulations. 
These quantum games are valued for their accessibility and intuitive engagement, especially for students with limited prior exposure to quantum mechanics \cite{seskirQuantumGamesInteractive2022corr}. 
Today, more than 300 quantum-related games exist \cite{ListQuantumGamescorr}, many of which are freely available and open source.

\subsubsection*{Hackathons} Quantum hackathons \cite{QmunityHackQThoncorr} have emerged as powerful environments for experiential learning. 
These events blend hands-on coding with peer collaboration and expert mentorship, enabling students to interact with real quantum devices and simulators. 
By fostering creativity and collaboration, they provide a motivating gateway into quantum computing for younger learners. 
Global initiatives such as QWorld \cite{qworld2025} and the Unitary Fund \cite{unitary2025} now host hackathons and coding camps specifically designed for high school and early university learners.

\subsubsection*{Summer schools} Quantum summer schools provide structured and immersive learning experiences during academic breaks. 
The Qubit by Qubit course developed in collaboration with MIT \cite{QubitxQubit} includes a dedicated summer program for high-school students, featuring live instruction, mentorship, and projects. 
These programs introduce quantum computing concepts in a scaffolded way that accommodates students from diverse academic backgrounds.

\subsection{Student-Led and Community Quantum Learning Initiatives}
A growing number of grassroots initiatives led by students and early-career researchers are redefining how quantum science is introduced to new learners. 
These efforts have emerged organically across the globe, often driven by individuals seeking to bridge the gap between advanced research and early-stage education. 
Many of these communities consist of high schoolers, undergraduates, and graduate students working collaboratively to produce open, engaging, and accessible content. Some examples include:
\begin{itemize}
    \item \emph{bqb Quantum Youth}~\cite{bqb2024}: A global student-led initiative co-founded by the authors, focused on producing beginner-friendly quantum science content and online events. 
    Spanning secondary school to PhD-level contributors, bqb emphasizes outreach to high school and undergraduate students using platforms already familiar to them, such as Instagram, YouTube, and TikTok.
    
    \item \emph{Quantum Universal Education (QUE)}~\cite{que2024}: One of the earliest student-led global networks in the space, QUE developed accessible quantum education resources and pioneered virtual events with a strong focus on inclusivity, particularly in underserved regions.
    
    \item \emph{Girls in Quantum}~\cite{giq2024}: An international nonprofit initiative led by young women, dedicated to empowering girls and underrepresented groups in quantum science through online talks, educational posts, and a supportive global network.
    
    \item \emph{QubitHub}~\cite{qubithub2024}: Based in Latin America, this student-academic collective produces multilingual infographics, webinars, and entry-level content to help learners build hands-on experience with quantum technologies.
\end{itemize}

These groups organize online events, publish tutorials, and create explainer videos aimed at making quantum concepts understandable and engaging for younger audiences, from middle school to early university levels. 
Their content is typically free and distributed through social media, open-source repositories, and virtual classrooms.

In addition to lowering entry barriers and providing mentorship, these initiatives create welcoming environments for students who may lack formal access to quantum education. 
At the same time, they offer valuable development opportunities for the students who lead them, helping them build skills in teaching, communication, leadership, and global collaboration.

Together, these decentralized, peer-led efforts highlight the power of student-driven, open education models to foster a more inclusive, motivated, and globally connected next generation of quantum learners.

\subsection{Addressing Regional Inequities with Open Resources}
A consistent theme in recent reports is the uneven global landscape of quantum education, with initiatives and advanced coursework largely concentrated in a small number of countries and elite institutions \cite{WorldHeadingQuantum2023corr,meyerDisparitiesAccessUS2024}. 
Regions with strong research ecosystems are more likely to be actively testing high-school quantum programs, while many others lag. 
In Latin America, for example, quantum education is still in its early stages, due to limited educational infrastructure and funding support \cite{tenjo-patinoQuantumComputingEducation2025}.
Few secondary schools in the region offer any related quantum topics, and opportunities often depend on isolated university outreach or online programs. 
The issue is not unique to this region: at a global level, high-school students have little to no exposure to quantum concepts in their curriculum. 

Open-source resources are currently the most accessible and widely used way to address global disparities in quantum education. % beyond the classroom.
Open online courses and freely available educational materials allow motivated students and educators worldwide to access high-quality quantum instruction, contingent on internet connectivity.

However, open access alone is insufficient to eliminate all forms of educational inequity.
Many schools lack the infrastructure, qualified instructors, or institutional awareness necessary to distribute and effectively utilize these resources.
Furthermore, language remains a major barrier, preventing many students from engaging with English-dominated content.
In response, language-specific efforts, particularly for Spanish-speaking communities, have begun to surface \cite{maldonado-romoQuantumComputingOnline2022corr}, though they remain modest in scale.
Additional challenges, such as unreliable internet connectivity and time zone differences, further hinder equitable participation in global online events.

Despite these constraints, as we have seen in this section, open-source efforts play a critical role in broadening access to quantum education, fostering international collaboration, and laying a foundation for more inclusive and sustainable quantum literacy.

\subsection{Combating Misinformation and Ensuring Quality} %TODO: Confirm these citations better
One of the major challenges of promoting education through open resources is the lack of consistent quality control and scope definition \cite{GuidelinesDevelopmentOpen}.
This problem is particularly acute in fast-evolving, interdisciplinary fields such as quantum technologies, where conceptual complexity and a rapidly shifting landscape make accurate communication difficult.

Quantum computing faces a well-documented hype problem \cite{meyerHowMediaHype2023corr,ezrattyMitigatingQuantumHype2022,smithiiiQuantumTechnologyHype2020corr}, which is often exacerbated by the widespread availability of unvetted online educational content.
In the absence of appropriate review mechanisms, these materials can disseminate misconceptions, exaggerate the capabilities of quantum devices, and create unrealistic expectations among learners.
This issue is particularly acute in outreach to high-school students, who may lack the conceptual background necessary to critically evaluate the accuracy of such content.
When educational resources are overhyped or inadequately explained, they risk generating disillusionment, reinforcing misconceptions, and ultimately eroding trust in scientific communication.

In contrast, the materials presented in \Cref{sec:testedmat} have been developed by domain experts and disseminated through peer-reviewed channels, ensuring both technical accuracy and pedagogical suitability.
Still, there remains a need for a more systematic approach to content verification.
One possible solution could be the establishment of an endorsing authority, such as a United Nations-backed body or an international consortium of experts, that certifies quantum education materials for public dissemination.
Such efforts would not only safeguard learners from misinformation but also contribute to building a reliable and trusted ecosystem for public engagement with quantum science \cite{zawacki-richterQualityAssuranceOpen2022corr}.
Comparable movements exist in adjacent technologies:
\begin{itemize} 
    \item \emph{Common Sense Education} \cite{CommonSenseEducation} evaluates and rates digital learning tools for schools; 
    \item \emph{IEEE REACH} \cite{IEEEREACH} develops curated historical and ethical engineering content for educators; 
    \item \emph{MIT’s RAISE} \cite{MITRAISEResponsible} initiative creates accessible, vetted AI education resources for youth.
\end{itemize}
While these are not formal certifying bodies, they demonstrate a growing recognition of the need for trustworthy, authoritative oversight in technology education.

It is worth noting that establishing such endorsing bodies raises important concerns: 
Who would determine their authority? 
How would they be funded? 
Could their gatekeeping role unintentionally discourage grassroots content creation? 
Would they introduce bias in what is deemed 'acceptable' educational material? 
Despite these challenges, we believe it is still worthwhile to explore such mechanisms. 
If well implemented, they can reduce misinformation and offer learners reliable, high-quality resources, which is especially valuable in complex fields like quantum technology.

\section{Conclusion}\label{sec:conclusion}
This paper has explored how quantum computing can be meaningfully integrated into high-school education, not as a standalone subject, but as a set of interconnected concepts embedded within existing STEM curricula. We discussed how mathematics, physics, and computer science modules vary across multiple national curricula, and we identified natural entry points where quantum concepts can be introduced. We also highlighted how project-based learning, open-source resources, and student-led initiatives can serve as powerful tools for broadening access to quantum education.

Despite growing global interest in developing a quantum-ready workforce, access to quantum education remains uneven, particularly for students in under-resourced regions. This disparity risks creating a “quantum divide,” in which only a narrow segment of students are prepared to participate in or benefit from the advances of this emerging field. We argued that accessible and reliable quantum open-source resources can help address this challenge.

We have also shown that high-school students are capable of engaging with complex quantum topics, especially when given access to hands-on activities, visual tools, and meaningful mentorship. Student projects, summer schools, games, and coding experiences offer complementary ways to deepen understanding beyond the traditional classroom. Equally important is the support provided to educators. Teacher training, professional development programs, and curated teaching materials are essential to empower instructors who may not have formal training in quantum science. We discussed how a modular approach, one that gradually integrates quantum computing into standard high-school topics, can be beneficial.

Looking ahead, building a globally inclusive quantum education ecosystem will require coordinated efforts among policymakers, researchers, educators, and the open-source community. Investment in localized and multilingual materials and equity-focused curriculum design will be key to ensuring that quantum education does not remain the privilege of a few.

Ultimately, preparing the next generation to engage confidently with quantum technologies is about cultivating thoughtful, scientifically literate citizens who can navigate and shape the technological challenges of the future. With modest but intentional steps, high-school education can become the foundation for a more equitable and sustainable quantum future.

\bibliography{combined}

\end{document}